\documentclass[11pt,a4paper]{article}
\pdfoutput=1
\usepackage{empheq}
\usepackage{cases}
\usepackage{a4wide}
\usepackage{amsfonts}
\usepackage{amssymb}
\usepackage{amsmath,bm}
\usepackage{amsmath}
\usepackage{graphicx}
\usepackage{mathtools}
\usepackage{booktabs}
\usepackage{array}
\usepackage{color}
\usepackage{ulem}
\usepackage[small,bf]{caption}
\usepackage{cite}
\usepackage[usenames,dvipsnames]{xcolor}
\usepackage{subfigure}
\usepackage{verbatim}

\allowdisplaybreaks

\numberwithin{equation}{section}

\newcounter{mysubequation}[equation]

\DeclarePairedDelimiter\bra{\langle}{\rvert}
\DeclarePairedDelimiter\ket{\lvert}{\rangle}
\DeclarePairedDelimiterX\braket[2]{\langle}{\rangle}{#1 \delimsize\vert #2}

\begin{document}
\begin{titlepage}

\begin{center}
{
\bf\LARGE Probing Cosmic Neutrino Background Charge\\[0.3em]
via Unconventional Interferometer
}
\\[8mm]
Chrisna~Setyo~Nugroho\footnote[1]{setyo13nugros@ntnu.edu.tw}  
\\[1mm]
\end{center}
\vspace*{0.50cm}

\centerline{\it Department of Physics, National Taiwan Normal University, Taipei 116, Taiwan}
\vspace*{1.20cm}

\begin{abstract}
\noindent
If neutrinos carry non-zero electric charge, they would interact 
directly
with photons. This would induce a phase shift along the photon
path in the optical experiment. We propose a novel idea to detect
this phase shift induced by
cosmic neutrino background (CNB) and the photon interaction using
laser interferometry experiment. We show that our setup can probe
the CNB neutrino charge in the order of $10^{-18} \,e-
10^{-17}\, e$. This is quite competitive with the existing upper bound on neutrino charge 
from both laboratory experiments and astrophysical observations.
\end{abstract}

\end{titlepage}
\setcounter{footnote}{0}

\section{Introduction}

The nature of the neutrino is one of the open problems in physics.
Although its interaction with matter is well understood within the
framework of the Standard Model of particle physics (SM), its non-zero mass, as implied by neutrino flavor oscillations, indicates the
necessity to go beyond the SM. Moreover, there is no symmetry that
forbids it to have non-vanishing electromagnetic moments~\cite{Ge:2022cib} as well as electric charge even in the SM~\cite{Foot:1990uf,Foot:1992ui}.
Indeed, the presence of neutrino charge is closely associated to the
nature of neutrino masses~\cite{Babu:1989ex,Babu:1989wn} (Dirac or Majorana)  as well as the charge
quantization. 
Several laboratory experiments such as low energy reactor neutrinos
and beam dump experiments as well as various astrophysical
observations have placed the upper limits on the neutrino
charge. 

On the other hand, the direct detection of relic neutrinos alias
CNB poses a great challenge to the
experimentalists due to their minuscule energies and cross sections.
There are several interesting proposals trying to address this problem
such as the use of the existing gravitational wave detectors~\cite{Domcke:2017aqj,Shergold:2021evs}, the absorption of CNB on
tritium~\cite{PTOLEMY:2018jst,Betts:2013uya,Banerjee:2023lrk}, the resonant
scattering between
CNB and very energetic cosmic neutrinos~\cite{Brdar:2022kpu}, cosmic
birefringence from CNB~\cite{Mohammadi:2021xoh}, as well as the
bremsstrahlung induced by the
CNB~\cite{Asteriadis:2022zmo}.
 
Assuming that the neutrinos carry non-zero electric charge as well as the asymmetric number of neutrinos with respect to the anti-neutrinos, we propose a novel strategy to detect the CNB neutrinos via laser interferometer.
When CNB neutrinos interact with the laser in one arm of the
interferometer, it would induce a phase shift on the laser to be
measured at the
output port. However, the existing exotic particle search proposals utilizing the laser interferometer employed at Gravitational Wave (GW) experiments~\cite{Tsuchida:2019hhc,Lee:2020dcd,Chen:2021apc,Ismail:2022ukp,Chen:2022abz,Lee:2022tsw,Seto:2004zu,Adams:2004pk,Riedel:2012ur,PhysRevLett.114.161301,Arvanitaki:2015iga,Stadnik:2015xbn,Branca:2016rez,Riedel:2016acj,Hall:2016usm,Jung:2017flg,Pierce:2018xmy,Morisaki:2018htj,Grote:2019uvn} are not
suitable for CNB neutrinos detection. In typical interferometer, both
of the interferometer arms would encounter the same amount of CNB neutrinos 
leading to zero phase shift in the photon path. We suggest to employ an
unconventional interferometer, widely known in quantum optics literature (see, for example \cite{Ou:2017}), which have a single arm as well as single photon mode. By utilizing
the squeezed vacuum state to probe the phase shift, we demonstrate
that our proposal is sensitive to probe the CNB neutrinos charge as low as the order of $10^{-18} \,e-
10^{-17}\, e$, with $e$ denotes the magnitude of electron charge.
 
The following is how the paper is organized: We discuss the interaction
between the non-relativistic CNB neutrinos and the photon in
Section~\ref{sec:interaction}. We introduce a phase measurement
based on unconventional
interferometer in Section~\ref{sec:phase} and further present the
projected sensitivities of our proposal 
in section~\ref{sec:result}.
Our summary and conclusions are presented in Section~\ref{sec:Summary}.

\section{CNB and Photon Interaction}
\label{sec:interaction}

Since the CNB neutrinos are non-relativistic, we start with  the
Hamiltonian describing the
interaction between non-relativistic charged particles and the
photon given by
\begin{align}
\label{eq:hamiltontot}
H = H_{P} + H_{R} + H_{I} \,,
\end{align}
where $H_{P}$, $H_{R}$, and $H_{I}$ stand for the free charged
particles, the free photon field, and the interaction between charged particles and the photon,
respectively. Their explicit expressions are~\cite{cohen:1987}
\begin{align}
\label{eq:hamilall}
H_{P} &= \sum_{\beta} \frac{\vec{p}^{2}_{\beta}}{2\, m_{\beta}} + V_{\text{Coulomb}}\,,\\
H_{R} &= \sum_{i} \hbar \omega_{i} \left( \hat{a}^{\dagger}_{i} \hat{a}_{i} + \frac{1}{2}\right)\,,\\
H_{I} &= H_{I1} + H_{I2}\,,\\
H_{I1} &= - \sum_{\beta} \frac{\text{q}_{\beta}}{m_{\beta}} \, \vec{p}_{\beta} \cdot \vec{A}(\vec{r}_{\beta})\,,\\
H_{I2} &= \sum_{\beta} \frac{\text{q}^{2}_{\beta}}{2\,m_{\beta}} \left[ \vec{A}(\vec{r}_{\beta})\right]^{2}\,.
\end{align}  
Here, $\vec{p}_{\beta}$, $m_{\beta}$, and $\text{q}_{\beta}$
denote the momentum, the mass, and the electric charge of the  $\beta$-th charged particle, respectively. The operator $\hat{a}_{i}$ ($\hat{a}^{\dagger}_{i}$) describes the annihilation (creation)
operator of the photon field for the i-th mode which obeys the
commutation relation $[\hat{a}_{i},\hat{a}^{\dagger}_{j} ] = \delta_{ij}$. In terms of these operators,  the photon field can be
written as~\cite{cohen:1987}
\begin{align}
\label{eq:photon}
\vec{A}(\vec{r}) = \sum_{i} \left[ \frac{\hbar}{2 \,\epsilon_{0}\, \omega_{i} L^{3}} \right]^{1/2} \left[\hat{a}_{i}\, \vec{\varepsilon}_{i} \, e^{\text{i} \vec{k}_{i} \cdot \vec{r}} + \hat{a}^{\dagger}_{i}\, \vec{\varepsilon}_{i} \, e^{-\text{i} \vec{k}_{i} \cdot \vec{r}} \right] \,,
\end{align}
where we have quantized the photon field $\vec{A}(\vec{r})$ in a box 
of volume $L^{3}$ with a normalization condition $\vec{k} \cdot \vec{L} = 2\pi\,n$ with $n$ denoting integer. Here,
the photon wave number and its the angular frequency satisfies $ \omega = |\vec{k}|\, c$. 

When the photon encounters the CNB neutrinos along its path, it
undergoes the phase shift $\delta$ due to the interaction between
them. Thus, the relevant Hamiltonian responsible for the
photon phase shift is $H_{I} = H_{I1} + H_{I2}$. The first term $H_{I1}$ is proportional
to $(\hat{a} + \hat{a}^{\dagger})$ which induces the energy
transition in a bound system. However, for a free particle system
such as CNB neutrinos, there is no
such a transition otherwise the energy conservation would be
violated.
Thus, one drops this term for free CNB neutrinos under
consideration.

On the other hand, the second term $H_{I2}$ is proportional to $(\hat{a} \hat{a} + \hat{a} \hat{a}^{\dagger} + \hat{a}^{\dagger} \hat{a} + \hat{a}^{\dagger} \hat{a}^{\dagger})$ which induces two
photons transition. For free CNB neutrinos, only the second and
the third term survive and one has
\begin{align}
\label{eq:HintF}
H_{I} \equiv \hat{H}_{\text{int}} &= \sum_{\beta} \frac{q^{2}_{\beta}}{2\,m_{\beta}} \, \left[ \frac{\hbar}{2 \,\epsilon_{0}\, \omega L^{3}} \right] 2\left(\hat{a}^{\dagger} \hat{a} + \frac{1}{2} \right)\nonumber \\
 &= \frac{\epsilon^{2}_{\nu} \, e^{2}}{m_{\nu}}\,\left[ \frac{\hbar\,\omega^{2}}{16\,\pi^{3} \,\epsilon_{0}\, c^{3}} \right] \left(\hat{a}^{\dagger} \hat{a} + \frac{1}{2} \right)\, N_{\nu}\,, 
\end{align} 
where we have assumed the CNB neutrinos have the same charge $q_{\beta} = \epsilon_{\nu}\,e$ and the same mass $m_{\beta} = m_{\nu}$ such that
the sum over CNB neutrinos is proportional to its total number $N_{\nu}$. This shows that the interaction between the photon and
CNB neutrinos is flavor blind since it depends on neutrino mass.
Note that the second line of Eq.\eqref{eq:HintF} is obtained by
substituting $L = 2\pi / k$ for $n = 1$ or single mode field. Such approximation is in
similar fashion with Jaynes-Cummings model in atomic transition~\cite{Jaynes:1963zz,Garrison:2008jh}. In practice, this is achievable by
incorporating frequency selective component such as Fabry-Perot etalon into the optical cavity which selects a single mode with the lowest loss~\cite{Fox:2006quantum}.  

\begin{figure}
	\centering
	\includegraphics[width=0.9\textwidth]{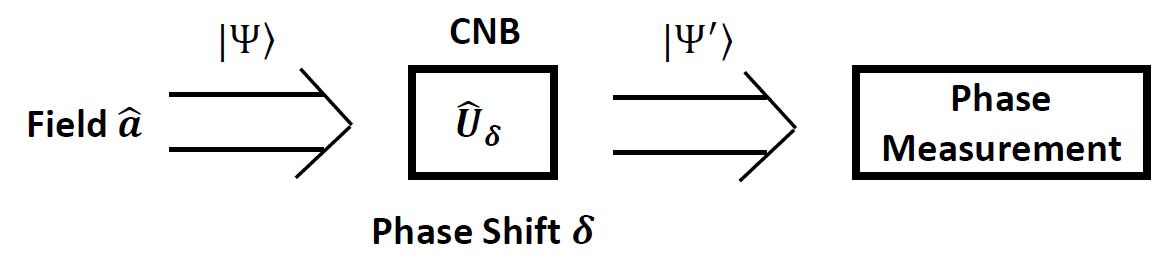}
	\caption{The phase shift $\delta$ on the optical field $\hat{a}$ changes the photon state from $\ket{\Psi}$ to $\ket{\Psi^{'}}$ due to photon-CNB neutrinos interaction.}
	\label{fig:interfero}
\end{figure}

We would
like to measure the phase shift
induced by the CNB neutrinos and the photon interaction through
the phase measurement scheme shown in
Fig.~\ref{fig:interfero}. The transformation of the photon state from $\ket{\Psi}$ to $\ket{\Psi^{'}}$ proceeds via the unitary operator $\hat{U}_{\delta}$~\cite{Kartner:1993,Ou:2017}
\begin{align}
\label{eq:psiP}
\ket{\Psi^{'}} &= \hat{U}_{\delta} \, \ket{\Psi} =e^{-\text{i}\,\hat{H}_{\text{int}}\text{t}/\hbar}\, \ket{\Psi}= e^{-\text{i}\,\hat{N}\delta}\,\ket{\Psi}\,,
\end{align}
where $ \hat{N} \equiv \hat{a}^{\dagger}\hat{a}$ denotes the photon number operator with the average value
$ N \equiv \left\langle \hat{N} \right\rangle = \left\langle \hat{a}^{\dagger}\hat{a} \right\rangle\gg 1$. 
However, due to the quantum nature the photon, the minimum
detectable phase shift for a given photon number $N$ is given by the Heisenberg
limit~\cite{Dirac:1927, Heitler:1954}
\begin{align}
\label{eq:HeisLimit}
\delta \geq \frac{1}{N}\,.
\end{align}
In optical laser experiment such as laser interferometer, 
the number of photon is of the
order of $10^{20}$ or larger which enables
us to detect a minuscule phase shift in the laboratory experiment.

\section{Phase Measurement Scheme}
\label{sec:phase}
Unfortunately, a direct photo-detection of the field $\hat{a}$ would
not allow us to extract its phase. The phase information can be
obtained via another unitary transformation of the field $\hat{a}$
which is a phase sensitive state. If we label the intial photon
state by $\ket{\Phi}$ and its final state after the phase shift by $\ket{\Phi^{'}}$, we can construct the phase sensitive state by a
unitary transformation $\hat{U}_{ps}$
\begin{align}
\label{eq:Ups}
\ket{\Psi} = \hat{U}_{ps}\,  \ket{\Phi},\,\, \ket{\Psi^{'}}= \hat{U}_{ps} \,\ket{\Phi^{'}}\,,
\end{align}
where the detection of $\ket{\Psi^{'}}$ would give a distinct
outcome from  $\ket{\Psi}$. Since both of these states are now
sensitive to the phase, we can detect the difference between them to
obtain the phase shift. In practice, one may employ the vacuum state as  
$\ket{\Psi}$ such that the phase measurement on it would yield null result. Thus, the detection of photon in $\ket{\Psi^{'}}$ implies the phase shift. 

\begin{figure}
	\centering
	\includegraphics[width=0.9\textwidth]{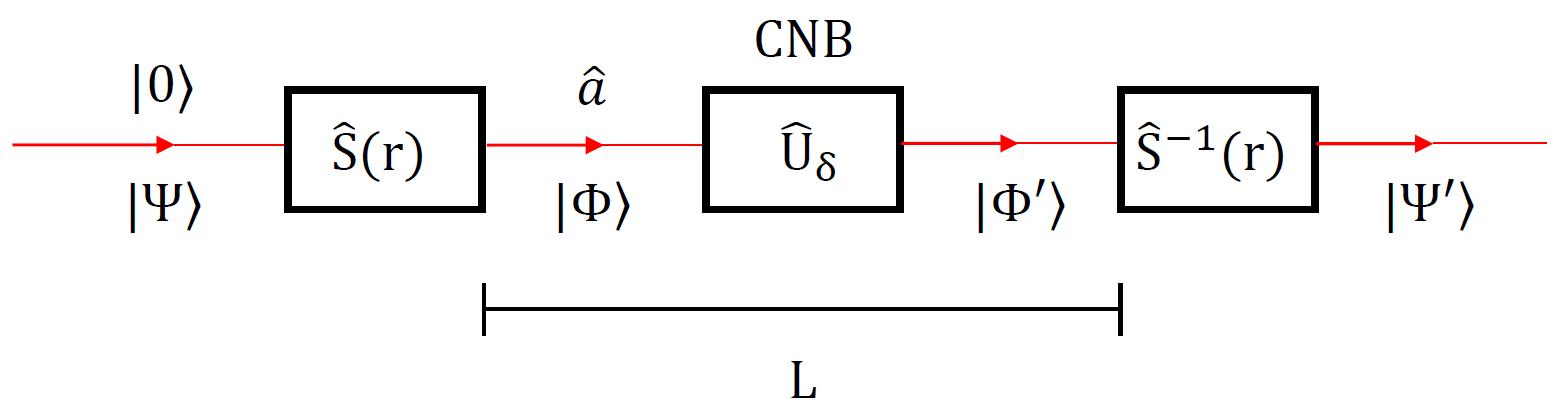}
	\caption{Single mode interferometer~\cite{Yurke:1986,Ou:2017} with vacuum state in its input. The presence of photons at the output indicates non-zero phase shift $\delta$ due to CNB neutrinos and photon interaction.}
	\label{fig:SMInterferometer}
\end{figure}

This scheme can be realized by using laser interferometer. However, in
conventional interferometer which consists of two arms such as
Michelson interferometer, the CNB neutrinos would interact with the
photon on both of arms resulting zero phase shift at the output. To
overcome this problem, we propose to employ unconventional
interferometer displayed in Fig.\ref{fig:SMInterferometer},  which was proposed for the first time in~\cite{Yurke:1986} and has been realized in the laboratories~\cite{Hudelist:2014,Manceau:2016esq,Liu:2018ahw,Xiao:2019,Ferreri:2020pxa}. Here, in
contrast to the conventional interferometer, the beam splitter is
replaced by the squeezing operator 
\begin{align}
\label{eq:defsqueeze}
\hat{S}(r) = e^{r (\hat{a}^{\dagger\,2} - \hat{a}^{2})/2}\,\,\, (r = \text{real}),
\end{align}
to form a single mode interferometer with the squeezed vacuum state $\ket{\Phi} = \hat{S}(r)\, \ket{0}$ acting as the probe of the phase shift $\delta$~\cite{Ou:2017}.  
The effects of the squeezing operator with squeezing parameter $r$
on the field operator $\hat{a}$ and its Hermitian conjugate $\hat{a}^{\dagger}$ are
\begin{align}
\label{eq:OpProperties}
\hat{S}^{\dagger}(r)\, \hat{a}\, \hat{S}(r) &= \hat{a}\, \text{cosh}\,r \, + \hat{a}^{\dagger} \, \text{sinh}\,r\,, \\
\hat{S}^{\dagger}(r)\, \hat{a}^{\dagger}\, \hat{S}(r) &= \hat{a}^{\dagger}\, \text{cosh}\,r \, + \hat{a} \, \text{sinh}\,r\,.
\end{align}
Thus, in our phase measurement scheme we have
\begin{align}
\label{eq:psState}
\hat{U}_{ps} = \hat{S}^{-1}(r) \,\,\, \text{and}\, \ket{\Psi} = \ket{0}\,,
\end{align}
where, in the absence of the phase shift, the output state would read $\ket{\Psi^{'}}_{\delta = 0} = \hat{S}^{-1}(r) \hat{S}(r) \ket{0} = \ket{0}$, see Fig.\ref{fig:SMInterferometer}. On the other hand, when there is a phase shift $\delta$, the output state becomes
\begin{align}
\label{eq:outState}
\ket{\Psi^{'}} = \hat{S}^{-1}(r)\, e^{-\text{i}\,\hat{a}^{\dagger} \hat{a}\, \delta}\, \hat{S}(r) \ket{0}\,.
\end{align}
Consequently, the output port of the interferometer would read the average photon number given by
\begin{align}
\label{eq:nout}
\bra{\Psi^{'}} \hat{a}^{\dagger} \hat{a} \ket{\Psi^{'}} = 4\left\langle \hat{N} \right\rangle  \left[ 1 + \left\langle \hat{N} \right\rangle \right]\, \text{sin}^{2} \delta\,,
\end{align}
where $\left\langle \hat{N} \right\rangle = \text{sinh}^{2} r$ denotes the
average photon number evaluated with respect to the state $\ket{\Phi}$.

Since no photon would be observed for zero phase shift, the presence
of photons at the output indicates a non-zero phase shift induced by
CNB-photon interaction. Taking
the noise at the output as one photon, the signal-to-noise ratio (SNR) reads~\cite{Ou:2017}
\begin{align}
\label{eq:SNR}
\text{SNR} = 4\left\langle \hat{N} \right\rangle  \left[ 1 + \left\langle \hat{N} \right\rangle \right]\, \text{sin}^{2} \delta\,.
\end{align}
The noise in our SNR calculation comes from the quantum nature of light. We neglect other noise components since they are sub-dominant and can easily be suppressed using current technology. In fact, the Heisenberg limit had been achieved in laboratories more than three decades ago \cite{Bondurant:1984,Grangier:1987,Xiao:1987} leaving only quantum noise relevant for our SNR computation.  
In the limit of $\left\langle N \right\rangle >> 1$ and $\delta << 1$,
one recovers the Heisenberg limit $\delta \sim 1/\left\langle N \right\rangle$ for SNR equals to 1.

\section{Results and Discussion}
\label{sec:result}
To obtain the sensitivity of our proposal, we require the SNR value
to be larger than one. We take the CNB neutrino density $n_{\nu} = 56\,\text{cm}^{-3}$~\cite{Ringwald:2004np}. Comparing Eq.~\eqref{eq:HintF} and Eq.~\eqref{eq:psiP}, the
phase shift experienced by the probe photon is
\begin{align}
\label{eq:delta}
\delta = \frac{\epsilon^{2}_{\nu} \, e^{2}}{m_{\nu}}\,\left[ \frac{\omega^{2}}{16\,\pi^{3} \,\epsilon_{0}\, c^{3}} \right] \, N_{\nu}\,t\,,
\end{align}
where $t$ is the interaction time between CNB neutrinos and the photons which is $\text{L}/c$.    
\begin{figure}
	\centering
	\includegraphics[width=14.0 cm]{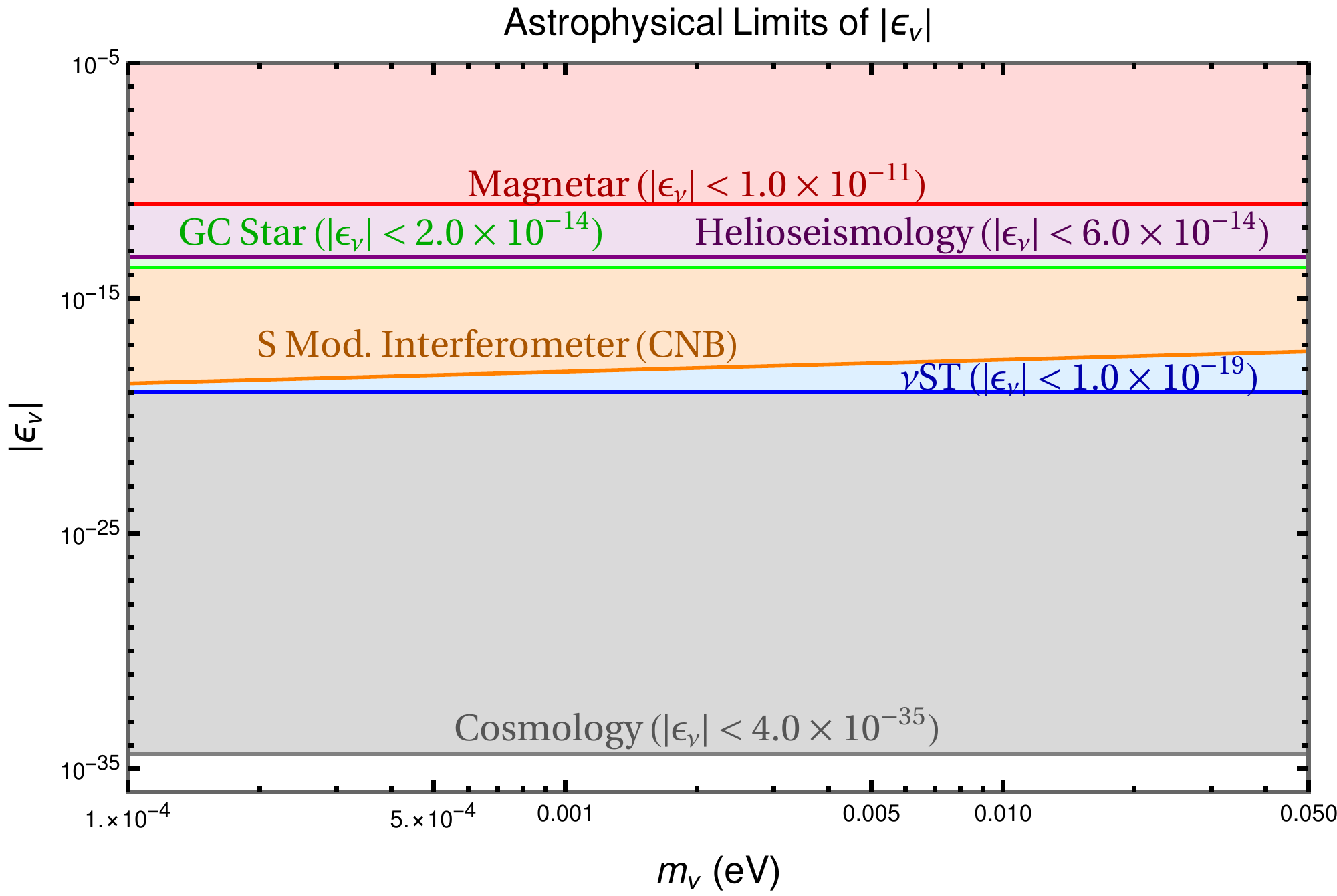}
	\caption{The projected sensitivity of single mode (S Mod.)
	interferometer with arm length L = 1 km and 1064 nm laser for flavor independent (flavor blind) neutrino. We take the phase sensing photon number $N = 10^{23}$~\cite{GammeVT-969:2007pci,Bahre:2013ywa,ALPS:2009des,Inada:2013tx}. The astrophysical bounds are also shown.} 
\label{fig:sensitivity}
\end{figure} 
Since the probe photon only interacts with CNB neutrinos along its path $\ell$, the
total CNB neutrinos number $N_{\nu}$ is obtained by integrating the
number of CNB neutrinos per unit length with respect to the total length
traversed by the photon
\begin{align}
\label{eq:ell}
N_{\nu} = \int^{\text{L}}_{0} d\ell\, \tilde{n}_{\nu}\,,
\end{align}
where L is the interferometer arm length and $\tilde{n}_{\nu} = n^{1/3}_{\nu}$ stands for the number of CNB neutrinos per unit length in $\text{cm}^{-1}$. Here, we set the arm length
to 1 km which is the typical length employed at GW experiment
such as LIGO~\cite{LIGOScientific:2016aoc} to reach an
unprecedented sensitivity.

The projected sensitivity of the single mode interferometer is
indicated by the light orange region in Fig.~\ref{fig:sensitivity} for flavor independent case. Here, we emphasize 
that flavor independent or flavor blind neutrino is the one with
definite mass eigenstate. As a result, it is possible to define the
effective neutrino charge $q^{\text{eff}}_{\nu}$ as superposition of
the three flavor components~\cite{AtzoriCorona:2022jeb,XMASS:2020zke} which is denoted as $|\epsilon_{\nu}|\,e$ in this paper.  
For
neutrino mass in the range between $10^{-4}$ eV to $0.05$ eV allowed by
the recent Particle Data Group~\cite{ParticleDataGroup:2022pth}, the
single mode interferometer can probe the magnitude of the
fractional charge $|\epsilon|$ of the CNB neutrinos
as low as $10^{-18} - 10^{-17}$. In addition, the upper bound on
$|\epsilon_{\nu}|$ from the neutron stars which have very large
magnetic field of the order $10^{14}$ G or larger (magnetar) is
$10^{-11}$~\cite{Das:2020egb} (light red area). Furthermore, the upper bounds of $
|\epsilon_{\nu}|$ from  helioseismology (light purple region) as well as
globular cluster (GC) stars (light green area) are $6.0 \times
10^{-14}$ 
and $2.0 \times 10^{-14}$~\cite{Raffelt:1999gv}, respectively. Finally, the effect of
neutrino charge on rotating magnetized neutron stars or the neutrino
star turning mechanism ($\nu$ST) puts the most stringent constraint
on $|\epsilon_{\nu}|$ to be less than $10^{-19}$~\cite{Studenikin:2012vi} (light blue area). All
these astrophysical constraints are model dependent and suffer from
the astrophysical uncertainties. We see that the projected sensitivity of our
proposal is quite
competitive with respect to these limits.
Furthermore, the upper limit on neutrino charge from cosmology  is $
|\epsilon_{\nu}| < 4.0 \times 10^{-35}$ \cite{Caprini:2003gz} (light gray area). This limit was derived by the following assumptions: the charge was generated during primordial phase transition, the charge was
distributed uniformly in the universe, the conductivity of the
universe is infinity, and the charge is a linear first order
perturbation in Friedmann-Robertson-Walker (FRW) model. These approximations are model dependent and suffer from
uncertainties just like the astrophysical limits. Our
proposal, on the other hand, based on the laser experiment in well controlled environment in the laboratory. This is in
similar motivation with the search of axion like particle (ALP) in laboratory by ALPS I and II collaboration \cite{Bahre:2013ywa} in the presence of more stringent limits from supernovae as well as star cooling \cite{Arias:2020tzl}.

\begin{figure}
	\centering
	\includegraphics[width=14.0 cm]{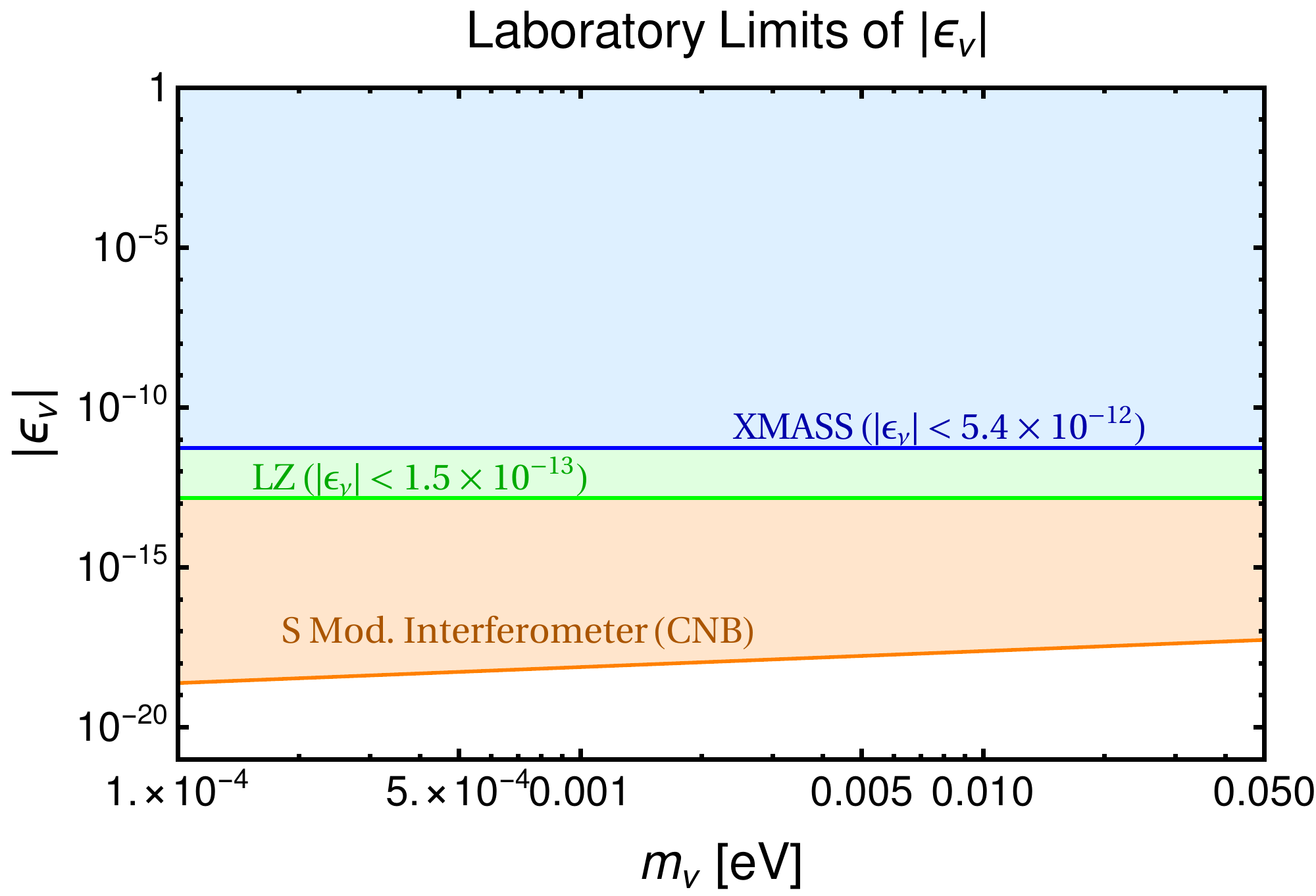}
	\caption{The projected sensitivity of single mode (S Mod.)
	interferometer with arm length L = 1 km and 1064 nm laser for flavor independent (flavor blind) neutrino along with the recent laboratory constraints. We take the phase sensing photon number $N = 10^{23}$~\cite{GammeVT-969:2007pci,Bahre:2013ywa,ALPS:2009des,Inada:2013tx}. } 
\label{fig:sensitivity2}
\end{figure}

On the other hand, there are several experiments that constraint the
upper limits on neutrino electric charge as shown in Fig.~\ref{fig:sensitivity2}. The XMASS collaboration places an upper limit
on millicharge neutrino as low as $ 5.4 \times 10^{-12} \,e$ at $90 \%$
confidence level~\cite{XMASS:2020zke} (light blue area).
Additionally, the strongest experimental constraint on neutrino
millicharge is $1.5 \times 10^{-13}\,e$ (light green area). This is obtained from LZ
experimental data using the Equivalent Photon Approximation (EPA)~\cite{AtzoriCorona:2022jeb}. Interestingly, this bound also depends
on the neutrino mass similar to our work. From the light orange
region in Fig.~\ref{fig:sensitivity2}, we see that our proposal is
several order magnitude more sensitive than the limits extracted from these experiments. 

Finally, let us discuss the existing neutrino charge limits based on its flavor. The strongest experimental limit on the first
generation neutrino charge is given by the combined data from beta decay $n \rightarrow p + e^{-} + \bar{\nu}_{e}$ and the non-neutrality of
matter~\cite{Raffelt:1999gv,Giunti:2014ixa}
\begin{align}
\label{eq:nuelim}
\epsilon_{\nu_{e}} = -0.6 \pm 3.2 \times 10^{-21}\,.
\end{align}
The upper limits
on the magnitude of the electron neutrino charge $|\epsilon_{\nu_{e}}|$ provided by TEXONO~\cite{Chen:2014dsa}, GEMMA collaboration~\cite{Beda:2009kx}, and CONUS experiment~\cite{CONUS:2022qbb} at
90$\%$ confident level are $2.1 \times 10^{-12}$, $1.5 \times 10^{-12}$, and $3.3 \times 10^{-12}$, respectively.
Moreover, the upper bound of $|\epsilon_{\nu_{e}}|$ from supernova SN
1987A is $10^{-16}$~\cite{Barbiellini:1987zz}.

For muon neutrino, the upperbound of $
|\epsilon_{\nu_{\mu}}|$ from the LSND experiment~\cite{LSND:2001akn} is $2.9 \times
10^{-9}$. DUNE collaboration restricts the muon neutrino millicharge in the interval $-9.3 \times 10^{-11} < \epsilon_{\nu_{\mu}} < 9.1 \times 10^{-11}$ at $90 \%$ confidence level~\cite{Mathur:2021trm}. However, the recent results from LZ collaboration~\cite{LZ:2022ufs,A:2022acy} give the most stringent limit on $
\epsilon_{\nu_{\mu}}$ in the range of $-6.2 \times 10^{-13} < \epsilon_{\nu_{\mu}} < 6.1 \times 10^{-13}$. Finally, the
upper bounds of $
|\epsilon_{\nu_{\tau}}|$ coming from the beam dump experiments
are $4.0 \times 10^{-4}$ (BEBC~\cite{Cooper-Sarkar:1991vsl,Babu:1993yh}) and $3.0 \times 10^{-4}$ (SLAC~\cite{Davidson:1991si}). In addition, the DONUT experiment~\cite{DONUT:2001zvi} puts more
stringent upper limit on $
|\epsilon_{\nu_{\tau}}|$ to be less than $3.9 \times 10^{-6}$. Furthermore, the LZ experiment~\cite{LZ:2022ufs,A:2022acy} provides the strongest constraint on $
\epsilon_{\nu_{\tau}}$ in the range of $-5.4 \times 10^{-13} < \epsilon_{\nu_{\tau}} < 5.2 \times 10^{-13}$.

\section{Summary and Conclusions}
\label{sec:Summary}

The existence of non-zero neutrino electric charges can be realized
in both SM and beyond the SM. It is closely related to the nature of
the neutrino itself whether it has a Dirac or Majorana mass
property.
There are several bounds on neutrino charge from both laboratory
experiments as well as astrophysical observations. On the other
hand, the direct search of relic neutrinos aka the CNB neutrinos
poses a serious difficulty to the experimentalists due to their
tiny energies and cross sections. 

Assuming that the CNB neutrinos have non-vanishing electric charges and there are more neutrinos than anti-neutrinos otherwise the net effect in the interferometer would cancel, 
we
propose a novel idea to detect them by utilizing unconventional
laser
interferometry experiment. Our aim is to detect the phase shift of
the laser induced by CNB neutrinos and photon interaction. We
estimate the SNR of this phase shift and found that the magnitude of
fractional charge of the CNB neutrinos $
|\epsilon_{\nu}|$ can be probed in the interval as low as $10^{-18} -
10^{-17}$. This is quite competitive with the existing astrophysical 
bounds for flavor universal case which are limited by astrophysical
uncertainties. Moreover, our proposal is several order magnitude more sensitive than the current laboratory limits. 
All in all, we
have shown that our proposed setup provides a promising venue to directly probe the CNB neutrino charge in the laboratory.

\section*{Acknowledgment}
\label{sec:Acknowledgment}
CSN is supported by the National Science and Technology Council (NSTC) of Taiwan under Grant No.MOST 111-2811-M-003-025- and 112-2811-M-003-004.


\begin{thebibliography}{99}

\bibitem{Ge:2022cib}
S.~F.~Ge and P.~Pasquini,
Phys. Lett. B \textbf{841}, 137911 (2023)
doi:10.1016/j.physletb.2023.137911
[arXiv:2206.11717 [hep-ph]].

\bibitem{Foot:1990uf}
R.~Foot, G.~C.~Joshi, H.~Lew and R.~R.~Volkas,
Mod. Phys. Lett. A \textbf{5}, 2721-2732 (1990)
doi:10.1142/S0217732390003176

\bibitem{Foot:1992ui}
R.~Foot, H.~Lew and R.~R.~Volkas,
J. Phys. G \textbf{19}, 361-372 (1993)
[erratum: J. Phys. G \textbf{19}, 1067 (1993)]
doi:10.1088/0954-3899/19/3/005
[arXiv:hep-ph/9209259 [hep-ph]].

\bibitem{Babu:1989ex}
K.~S.~Babu and R.~N.~Mohapatra,
Phys. Rev. D \textbf{41}, 271 (1990)
doi:10.1103/PhysRevD.41.271

\bibitem{Babu:1989wn}
K.~S.~Babu and R.~N.~Mohapatra,
Phys. Rev. Lett. \textbf{63}, 228 (1989)
doi:10.1103/PhysRevLett.63.228

\bibitem{Domcke:2017aqj}
V.~Domcke and M.~Spinrath,
JCAP \textbf{06}, 055 (2017)
doi:10.1088/1475-7516/2017/06/055
[arXiv:1703.08629 [astro-ph.CO]].

\bibitem{Shergold:2021evs}
J.~D.~Shergold,
JCAP \textbf{11}, no.11, 052 (2021)
doi:10.1088/1475-7516/2021/11/052
[arXiv:2109.07482 [hep-ph]].

\bibitem{PTOLEMY:2018jst}
E.~Baracchini \textit{et al.} [PTOLEMY],
[arXiv:1808.01892 [physics.ins-det]].

\bibitem{Betts:2013uya}
S.~Betts, W.~R.~Blanchard, R.~H.~Carnevale, C.~Chang, C.~Chen, S.~Chidzik, L.~Ciebiera, P.~Cloessner, A.~Cocco and A.~Cohen, \textit{et al.}
[arXiv:1307.4738 [astro-ph.IM]].

\bibitem{Banerjee:2023lrk}
I.~K.~Banerjee, U.~K.~Dey, N.~Nath and S.~S.~Shariff,
[arXiv:2304.02505 [hep-ph]].

\bibitem{Brdar:2022kpu}
V.~Brdar, P.~S.~B.~Dev, R.~Plestid and A.~Soni,
Phys. Lett. B \textbf{833}, 137358 (2022)
doi:10.1016/j.physletb.2022.137358
[arXiv:2207.02860 [hep-ph]].

\bibitem{Mohammadi:2021xoh}
R.~Mohammadi, J.~Khodagholizadeh, M.~Sadegh, A.~Vahedi and S.~s.~Xue,
[arXiv:2109.00152 [hep-ph]].

\bibitem{Asteriadis:2022zmo}
K.~Asteriadis, A.~Q.~Trivi\~no and M.~Spinrath,
[arXiv:2208.01207 [hep-ph]].


\bibitem{Tsuchida:2019hhc}
S.~Tsuchida, N.~Kanda, Y.~Itoh and M.~Mori,
Phys. Rev. D \textbf{101} (2020) no.2, 023005
[arXiv:1909.00654 [astro-ph.HE]].

\bibitem{Lee:2020dcd}
C.~H.~Lee, C.~S.~Nugroho and M.~Spinrath,
Eur. Phys. J. C \textbf{80}, no.12, 1125 (2020)
doi:10.1140/epjc/s10052-020-08692-3
[arXiv:2007.07908 [hep-ph]].

\bibitem{Chen:2021apc}
C.~R.~Chen and C.~S.~Nugroho,
Phys. Rev. D \textbf{105}, no.8, 083001 (2022)
doi:10.1103/PhysRevD.105.083001
[arXiv:2111.11014 [hep-ph]].

\bibitem{Ismail:2022ukp}
M.~A.~Ismail, C.~S.~Nugroho and H.~T.~K.~Wong,
Phys. Rev. D \textbf{107}, no.8, 082002 (2023)
doi:10.1103/PhysRevD.107.082002
[arXiv:2211.13384 [hep-ph]].

\bibitem{Chen:2022abz}
C.~R.~Chen, B.~H.~Nhung and C.~S.~Nugroho,
Phys. Lett. B \textbf{839}, 137764 (2023)
doi:10.1016/j.physletb.2023.137764
[arXiv:2212.13017 [hep-ph]].

\bibitem{Lee:2022tsw}
C.~H.~Lee, R.~Primulando and M.~Spinrath,
Phys. Rev. D \textbf{107}, no.3, 035029 (2023)
doi:10.1103/PhysRevD.107.035029
[arXiv:2208.06232 [hep-ph]].

\bibitem{Seto:2004zu}
N.~Seto and A.~Cooray,
Phys. Rev. D \textbf{70} (2004), 063512
[arXiv:astro-ph/0405216 [astro-ph]].

\bibitem{Adams:2004pk}
A.~W.~Adams and J.~S.~Bloom,
[arXiv:astro-ph/0405266 [astro-ph]].

\bibitem{Riedel:2012ur}
C.~J.~Riedel,
Phys.\ Rev.\ D {\bf 88} (2013) no.11,  116005
[arXiv:1212.3061 [quant-ph]].

\bibitem{PhysRevLett.114.161301}
Y.~V.~Stadnik and V.~V.~Flambaum,
Phys.\ Rev.\ Lett.\  {\bf 114} (2015) 161301
[arXiv:1412.7801 [hep-ph]].

\bibitem{Arvanitaki:2015iga}
A.~Arvanitaki, S.~Dimopoulos and K.~Van Tilburg,
Phys.\ Rev.\ Lett.\  {\bf 116} (2016) no.3,  031102
[arXiv:1508.01798 [hep-ph]].

\bibitem{Stadnik:2015xbn}
Y.~V.~Stadnik and V.~V.~Flambaum,
Phys.\ Rev.\ A {\bf 93} (2016) no.6,  063630
[arXiv:1511.00447 [physics.atom-ph]].

\bibitem{Branca:2016rez}
A.~Branca {\it et al.},
Phys.\ Rev.\ Lett.\  {\bf 118} (2017) no.2,  021302
[arXiv:1607.07327 [hep-ex]].

\bibitem{Riedel:2016acj}
C.~J.~Riedel and I.~Yavin,
Phys.\ Rev.\ D {\bf 96} (2017) no.2,  023007
[arXiv:1609.04145 [quant-ph]].

\bibitem{Hall:2016usm}
E.~D.~Hall, R.~X.~Adhikari, V.~V.~Frolov, H.~Müller, and M.~Pospelov,
Phys. Rev. D \textbf{98} (2018) no.8, 083019
[arXiv:1605.01103 [gr-qc]].

\bibitem{Jung:2017flg}
S.~Jung and C.~S.~Shin,
Phys.\ Rev.\ Lett.\  {\bf 122} (2019) no.4,  041103
[arXiv:1712.01396 [astro-ph.CO]].

\bibitem{Pierce:2018xmy}
A.~Pierce, K.~Riles and Y.~Zhao,
Phys.\ Rev.\ Lett.\  {\bf 121} (2018) no.6,  061102
[arXiv:1801.10161 [hep-ph]].


\bibitem{Morisaki:2018htj}
S.~Morisaki and T.~Suyama,
Phys.\ Rev.\ D {\bf 100} (2019) no.12,  123512
[arXiv:1811.05003 [hep-ph]].

\bibitem{Grote:2019uvn}
H.~Grote and Y.~V.~Stadnik,
Phys.\ Rev.\ Research.\  {\bf 1} (2019) 033187
[arXiv:1906.06193 [astro-ph.IM]].

\bibitem{Ou:2017}
Z.~Y.~J.~Ou,
Quantum Optics for Experimentalists,
World Scientific (2017)


\bibitem{cohen:1987}
C.~C.~Tannoudji, J.~D.~Roc, and G.~Grynberg,
Photons and Atoms Introduction to Quantum Electrodynamics,
John Wiley and Sons (1987),
doi:https://doi.org/10.1002/9783527618422.ch3

\bibitem{Jaynes:1963zz}
E.~T.~Jaynes and F.~W.~Cummings,
IEEE Proc. \textbf{51}, 89-109 (1963)
doi:10.1109/PROC.1963.1664

\bibitem{Garrison:2008jh}
J.~C.~Garrison and R.~Y.~Chiao,
Quantum Optics,
Oxford University Press (2008),
doi:https://doi.org/10.1093/acprof:oso/9780198508861.001.0001

\bibitem{Fox:2006quantum}
M.~Fox,
Quantum Optics: An Introduction,
Oxford University Press (2006)



\bibitem{Kartner:1993}
F.~X.~Kartner and H.~A.~Haus,
Phys.~Rev.~A \textbf{65}, p. 4585 (1993).




\bibitem{Dirac:1927}
P.~A.~M.~Dirac
Proc. R. Soc. London Ser. A \textbf{114}, p.243 (1927).

\bibitem{Heitler:1954}
W.~Heitler, The Quantum Theory of Radiation, 3rd edn., Oxford University Press, London (1954).

\bibitem{Yurke:1986}
B.~Yurke, S.~L.~McCall, and J.~R.~Klauder
Phys. Rev. A \textbf{33}, p.4033 (1986)

\bibitem{Hudelist:2014}
Hudelist, F., Kong, J., Liu, C. et al. Quantum metrology with parametric amplifier-based photon correlation interferometers, Nat Commun 5, 3049 (2014). https://doi.org/10.1038/ncomms4049.


\bibitem{Manceau:2016esq}
M.~Manceau, F.~Khalili and M.~Chekhova,
New J. Phys. \textbf{19}, no.1, 013014 (2017)
doi:10.1088/1367-2630/aa53d1
[arXiv:1607.07332 [quant-ph]].


\bibitem{Liu:2018ahw}
S.~Liu, Y.~Lou, J.~Xin and J.~Jing,
Phys. Rev. Applied \textbf{10}, no.6, 064046 (2018)
doi:10.1103/PhysRevApplied.10.064046


\bibitem{Xiao:2019}
Xiao Xiao et al. “Enhancement of Sensitiv-
ity by Initial Phase Matching in SU(1,1)
Interferometers”. In: Communications in
Theoretical Physics 71.1 (Jan. 2019),
p. 037. doi: 10.1088/0253-6102/71/1/37.

\bibitem{Ferreri:2020pxa}
A.~Ferreri, M.~Santandrea, M.~Stefszky, K.~H.~Luo, H.~Herrmann, C.~Silberhorn and P.~R.~Sharapova,
Quantum \textbf{5}, 461 (2021)
doi:10.22331/q-2021-05-27-461
[arXiv:2012.03751 [quant-ph]].


\bibitem{Bondurant:1984}
R.~S.~Bondurant and J.~H.~Shapiro
Phys.~Rev.~D \textbf{30}, p. 2548 (1984).

\bibitem{Grangier:1987}
P.~Grangier, R.~E.~Slusher, B.~Yurke, and A.~Laporta,
Phys. Rev. Lett. \textbf{59}, p.2153 (1987).

\bibitem{Xiao:1987}
M.~Xiao, L.~A.~Wu, and H.~J.~Kimble
Phys. Rev. Lett. \textbf{59}, p.278 (1987).


\bibitem{Ringwald:2004np}
A.~Ringwald and Y.~Y.~Y.~Wong,
JCAP \textbf{12}, 005 (2004)
doi:10.1088/1475-7516/2004/12/005
[arXiv:hep-ph/0408241 [hep-ph]].

\bibitem{LIGOScientific:2016aoc}
B.~P.~Abbott \textit{et al.} [LIGO Scientific and Virgo],
Phys. Rev. Lett. \textbf{116}, no.6, 061102 (2016)
doi:10.1103/PhysRevLett.116.061102
[arXiv:1602.03837 [gr-qc]].


\bibitem{GammeVT-969:2007pci}
A.~S.~Chou \textit{et al.} [GammeV (T-969)],
Phys. Rev. Lett. \textbf{100}, 080402 (2008)
doi:10.1103/PhysRevLett.100.080402
[arXiv:0710.3783 [hep-ex]].

\bibitem{Bahre:2013ywa}
R.~B\"ahre, B.~D\"obrich, J.~Dreyling-Eschweiler, S.~Ghazaryan, R.~Hodajerdi, D.~Horns, F.~Januschek, E.~A.~Knabbe, A.~Lindner and D.~Notz, \textit{et al.}
JINST \textbf{8}, T09001 (2013)
doi:10.1088/1748-0221/8/09/T09001
[arXiv:1302.5647 [physics.ins-det]].

\bibitem{ALPS:2009des}
K.~Ehret \textit{et al.} [ALPS],
Nucl. Instrum. Meth. A \textbf{612}, 83-96 (2009)
doi:10.1016/j.nima.2009.10.102
[arXiv:0905.4159 [physics.ins-det]].

\bibitem{Inada:2013tx}
T.~Inada, T.~Namba, S.~Asai, T.~Kobayashi, Y.~Tanaka, K.~Tamasaku, K.~Sawada and T.~Ishikawa,
Phys. Lett. B \textbf{722}, 301-304 (2013)
doi:10.1016/j.physletb.2013.04.033
[arXiv:1301.6557 [physics.ins-det]].

\bibitem{AtzoriCorona:2022jeb}
M.~Atzori Corona, W.~M.~Bonivento, M.~Cadeddu, N.~Cargioli and F.~Dordei,
Phys. Rev. D \textbf{107}, no.5, 053001 (2023)
doi:10.1103/PhysRevD.107.053001
[arXiv:2207.05036 [hep-ph]].

\bibitem{XMASS:2020zke}
K.~Abe \textit{et al.} [XMASS],
Phys. Lett. B \textbf{809}, 135741 (2020)
doi:10.1016/j.physletb.2020.135741
[arXiv:2005.11891 [hep-ex]].


\bibitem{ParticleDataGroup:2022pth}
R.~L.~Workman \textit{et al.} [Particle Data Group],
PTEP \textbf{2022}, 083C01 (2022)
doi:10.1093/ptep/ptac097


\bibitem{Das:2020egb}
A.~Das, D.~Ghosh, C.~Giunti and A.~Thalapillil,
Phys. Rev. D \textbf{102}, no.11, 115009 (2020)
doi:10.1103/PhysRevD.102.115009
[arXiv:2005.12304 [hep-ph]].


\bibitem{Raffelt:1999gv}
G.~G.~Raffelt,
Phys. Rept. \textbf{320}, 319-327 (1999)
doi:10.1016/S0370-1573(99)00074-5

\bibitem{Studenikin:2012vi}
A.~I.~Studenikin and I.~Tokarev,
Nucl. Phys. B \textbf{884}, 396-407 (2014)
doi:10.1016/j.nuclphysb.2014.04.026
[arXiv:1209.3245 [hep-ph]].

\bibitem{Caprini:2003gz}
C.~Caprini, S.~Biller and P.~G.~Ferreira,
JCAP \textbf{02} (2005), 006
doi:10.1088/1475-7516/2005/02/006
[arXiv:hep-ph/0310066 [hep-ph]].


\bibitem{Arias:2020tzl}
P.~Arias, A.~Arza, J.~Jaeckel and D.~Vargas-Arancibia,
JCAP \textbf{05}, 070 (2021)
doi:10.1088/1475-7516/2021/05/070
[arXiv:2007.12585 [hep-ph]].



\bibitem{Giunti:2014ixa}
C.~Giunti and A.~Studenikin,
Rev. Mod. Phys. \textbf{87}, 531 (2015)
doi:10.1103/RevModPhys.87.531
[arXiv:1403.6344 [hep-ph]].

\bibitem{Chen:2014dsa}
J.~W.~Chen, H.~C.~Chi, H.~B.~Li, C.~P.~Liu, L.~Singh, H.~T.~Wong, C.~L.~Wu and C.~P.~Wu,
Phys. Rev. D \textbf{90}, no.1, 011301 (2014)
doi:10.1103/PhysRevD.90.011301
[arXiv:1405.7168 [hep-ph]].

\bibitem{Beda:2009kx}
A.~G.~Beda, E.~V.~Demidova, A.~S.~Starostin, V.~B.~Brudanin, V.~G.~Egorov, D.~V.~Medvedev, M.~V.~Shirchenko and T.~Vylov,
Phys. Part. Nucl. Lett. \textbf{7}, 406-409 (2010)
doi:10.1134/S1547477110060063
[arXiv:0906.1926 [hep-ex]].

\bibitem{CONUS:2022qbb}
H.~Bonet \textit{et al.} [CONUS],
Eur. Phys. J. C \textbf{82}, no.9, 813 (2022)
doi:10.1140/epjc/s10052-022-10722-1
[arXiv:2201.12257 [hep-ex]].

\bibitem{Barbiellini:1987zz}
G.~Barbiellini and G.~Cocconi,
Nature \textbf{329}, 21-22 (1987)
doi:10.1038/329021b0

\bibitem{LSND:2001akn}
L.~B.~Auerbach \textit{et al.} [LSND],
Phys. Rev. D \textbf{63}, 112001 (2001)
doi:10.1103/PhysRevD.63.112001
[arXiv:hep-ex/0101039 [hep-ex]].

\bibitem{Mathur:2021trm}
V.~Mathur, I.~M.~Shoemaker and Z.~Tabrizi,
JHEP \textbf{10}, 041 (2022)
doi:10.1007/JHEP10(2022)041
[arXiv:2111.14884 [hep-ph]].

\bibitem{LZ:2022ufs}
J.~Aalbers \textit{et al.} [LZ],
[arXiv:2207.03764 [hep-ex]].

\bibitem{A:2022acy}
S.~K.~A., A.~Majumdar, D.~K.~Papoulias, H.~Prajapati and R.~Srivastava,
Phys. Lett. B \textbf{839}, 137742 (2023)
doi:10.1016/j.physletb.2023.137742
[arXiv:2208.06415 [hep-ph]].

\bibitem{Cooper-Sarkar:1991vsl}
A.~M.~Cooper-Sarkar, S.~Sarkar, J.~Guy, W.~Venus, P.~O.~Hulth and K.~Hultqvist,
Phys. Lett. B \textbf{280}, 153-158 (1992)
doi:10.1016/0370-2693(92)90789-7

\bibitem{Babu:1993yh}
K.~S.~Babu, T.~M.~Gould and I.~Z.~Rothstein,
Phys. Lett. B \textbf{321}, 140-144 (1994)
doi:10.1016/0370-2693(94)90340-9
[arXiv:hep-ph/9310349 [hep-ph]].

\bibitem{Davidson:1991si}
S.~Davidson, B.~Campbell and D.~C.~Bailey,
Phys. Rev. D \textbf{43}, 2314-2321 (1991)
doi:10.1103/PhysRevD.43.2314

\bibitem{DONUT:2001zvi}
R.~Schwienhorst \textit{et al.} [DONUT],
Phys. Lett. B \textbf{513}, 23-29 (2001)
doi:10.1016/S0370-2693(01)00746-8
[arXiv:hep-ex/0102026 [hep-ex]].






\end{thebibliography}
\end{document}